# Electronic Authority Variation

M.N.Doja

CSE Department
Jamia Millia Islamia
New Delhi, India
ndoja@yahoo.com

Dharmender Saini

CSE Department
Jamia Millia Islamia
New Delhi, India
dsaini77@yahoo.com

*Abstract*— **when a person joins in an organization, he becomes authorize to take some decisions on behalf of that organization; means he is given some authority to exercise. After some time, on the basis of his performance in the organization, he is given promotion and he becomes eligible to exercise to some higher authorities. And further, he may get some higher promotion or he may leave the organization. So, during his stay in the organization, the authority of that person varies from the time he joins the organization until he/she leaves the organization. This paper presents the variation in authorities of a person in the organization. The method implements the queuing model to analyze the various people in the queue of their promotion and looks at various parameters like average waiting time etc.**

*Keywords- Authority: Authority Variation: Authority Level*

## I.    INTRODUCTION

The problem of authorization was raised in 1990 by Fischer [1] for he confirmation of the originality of source. Russell [2] in 1994 described the problem in detail and suggested various options available to the receiver. He suggested some basic principles of authorization at source like auditing by receiver, trusted   third party originator, and self audit. He further categorized authorization in two parts i.e. person based authorization and rule based authorization. Person based authorization uses digital signatures and certificates, where as a rule based authorization is based on rules provided to the receiver for verification of authorization. Thomas woo [3] in 1998 suggested the design of distributed authorization service which parallels existing authentication services for distributed systems. In 2000 Michiharu and Santoshi [4] presented xml document security based on provisional authorization. They suggested an xml access control language (XACL) that integrates security features such as authorization, non-repudiation, confidentiality, and an audit trail for xml documents. During the period of 1996 to 2005 various types of authorization and its application like [5, 6, 7] were suggested.  In 2005 [8] Burrows presented a method for XSL/XML based authorization rules policy implementation through filing a patent in united state patent office.  he implemented XSL/XML based authorization rules policy on a given set of data and used an authorization rules engine which uses authorization rules defined in XSL to operate on access decision information (ADI) provided by the user. Inside the authorization rules engine, a boolean authorization rules mechanism is implemented to constrain the XSL processor to arrive at a boolean authorization decision. When a person joins in an organization, he becomes authorize to take some decisions on behalf of that organization; means he is given some authority to exercise. After some time, on the basis of his performance in the organization, he is given promotion to some higher level and he becomes eligible to exercise to some higher authorities. And further, he may get some higher promotion or he may leave the organization. So, during his stay in the organization, the authority of that person varies from the time he joins the organization until he/she leaves the organization and also he remain in the queue [10, 11, 12] for next position. This paper presents the variation in authorities of a person in the organization. As soon as the person gets the promotion his/her authority database is updated to reflect the current authorities. The method implements the queuing model to analyze the various people in the queue of their promotion and looks at various parameters like average waiting time etc.

This paper is organized in four parts. Part 1 presents introduction to the problem addressed, Part II explains the Queuing Theory basics, Part III presents the Queuing Model implementation for our scheme and Part IV presents XML Policy for the User is used in this system. Part V presents Authority Variation when a person moves from one level of the queue to other level. Part VI presents conclusion and Part VII presents application and future scope.

## II. QUEUING THEORY BASICS

Queuing theory [10, 11] is a mathematical concept which is used the application study its various application in technology and all other related areas. We have used this concept to study people in organization are basically a queue of various points for example they may be in promotion queues. When they enter in the organization they are in the queue. When they are inside in the organization they are in the queue of promotion. Here we are studying the case when they are inside the organization and they are in the promotion queue. For example employees from level one L1 gets promotion to higher level two L2 and then higher and so on. But for simplicity, here we have taken only three levels i.e. L1, L2, and L3.

So, just for brief introduction to this concept the brief introduction of this concept is presented.

The three basic terms in queuing theory are customers, queues, and servers.

### A. Customers

Customers are generated by an input source. The customers are generated according to a statistical distribution and the distribution describes their interarrival times, i.e the times between arrivals of customers. The customers join a queue. In our system customers are person joining the organization.

### B. Server (Service Mechanism)

Customers are selected for service by the server at various times. The rule on the basis of which the customers are selected is called the queue discipline. The head of the queue is the customer who arrived in the queue first and tale, a person who is in the last. In our system the server is the organization authorities.

### C. Input Source

The input source is a population of individuals, and as such is called the calling population. The calling population has a size, which is the number of potential customers to the system. The size can either be finite or infinite. The input source in our system is the process which supplies person to the organization department fro example Human Resource Process.

### D. Queue

Queues are either infinite or finite. If a queue is finite, it holds a limited number of customers. The amount of time a customer waits in the queue is called the queuing time. The number of customers who arrive from the calling population and join the queue in a given period of time is modeled by a statistical distribution. In our system, we have taken the queue of people, waiting for their promotion or their authority to upgrade.

### E. Queue Discipline

The queue discipline is a rule through which customers are selected from the queue for processing by servers. For example, first-come-first-served (FCFS), where the customers are processed in the order they arrived in the queue. Most queuing models assume FCFS as the queue discipline. We have also assumed the same approach. In our system the queue discipline is the rule on the basis of which the promotion of employees occur.

### F. Basic Notations [10]

$\lambda n$ : Mean arrival rate of new customers when n customers are in the system.

$\mu n$ : Mean service rate (expected number of customers completing service per unit time) when n customers are in the system.

$P(i)$ : Probability of exactly i customers in queueing system.

$L$ : Expected number of customers in the queueing system.

$LS$ : Average waiting in the system..

## III. QUEUING MODEL IMPLEMENTATION

When a person joins in an organization, he becomes authorize to take some decisions on behalf of that organization; means he is given some authority to exercise. After some time, on the basis of his performance in the organization, he is given promotion and he becomes eligible to exercise to some higher authorities. And further, he may get some higher promotion or he may leave the organization. So, during his stay in the organization, the authority of that person varies from the time he joins the organization until he/she leaves the organization. This paper presents the variation in authorities of a person in the organization. The method implements the queuing model to analyze the various people in the queue of their promotion and looks at various parameters like average waiting time etc.

### A. Assumptions

- Let's take three queues implementing three level (for simplicity we have taken only three level hierarchy) various level of the employees in the organization

L1: Contain employees of the organization who just join,

L2: Contain employees at the second level after their promotion from first level i.e, first queue L1, and

L3: contains employees at next higher level promoted from previous level. i.e L2.

- For every level, Decide $\lambda n$ rate at which person are coming into the system and $\mu n$ denotes the rate at which person are going out of the system.

The below mentioned Figure1 shows various levels in the system where employees happens to be in the queue of promption.

Figure1. Levels of Various Queues

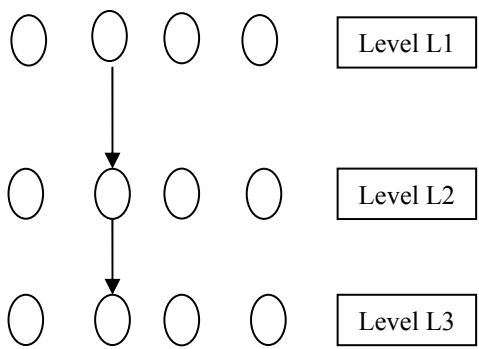

### IV. XML POLICY FOR THE USER

The xml policy for the user contains information about the  user who is signing the document the policy may contain information like  user identification, his hierarchy or designation, his  authorities  for whether he has the power of signing this document or not if yes in what capacity.

There is a database maintained in the organization of xml policies for verification of proper authority for the person who is exercising his/her the document. The structure of database is as shown in the table 1. The first column table describes the Employees identification number given by the organization and the second column describes the XML policies associated with the person for verifying his/her authorities.

Table 1 An Authority database

| Employees ID | XML policies |
|---|---|
| 0 | |
| 1 | |

The example of authorization policies [9] can be describing a person signing capabilities can be

```
<Policy>
<user>
 <name>smith</name>
 <id>1</id>
 <designation>manager</designation>
 <signing_limit>1000</ signing_limit>
</user>
</Policy>
```

### V. AUTHORITY VARIATION

When the person in an organization move from one level to other level their authority changes. For example, authority to review document, authority to sign document, authority to review people performance etc. The authority database for a person who got promotion should be updated. So, the table 1 records all the changes in person authorities and when person exercise his/her authority this table is referred and the policy for that person is verified according to the following stylesheet code.

```
<?xml version = "1.0"?>
<xsl:stylesheet version ="1.0" xmlns:xsl="www.w3.org/1999/XSLT/Transform">
<xsl:output method='html'/>
<xsl:template match="/">
```

```
<html>

<body>

<xsl: select = "//" />

<xsl:if test= "user/name='smith'">

<xsl:if test= "user/id='1'">

<xsl:if test= "user/designation='manager'">

<xsl:if test= "user/designation='1000'">

<table border="1">

<tr bgcolor = "#1acd31"><td>

!Access allowed

</td></tr></table>

</xsl:if

</body>

</html>

</xsl:template>

</xsl:stylesheet>
```

OUTPUT: **True**

The above result 'True' means that the person has exercised the right authority.

## VI.   CONCLUSION

We have presented the variation in authorities of a person in the organization as he moves from one higher level to other higher level means from one queue to another queue of promotion. We have implemented the queuing model to analyze the various people in the queue of their promotion and looked at various parameters like average waiting time. The method also implements the authority policy as a database of XML policies so that they can be referred at the time of taking decision about the authority of en employee.

## VII.   APPLICATION AND FUTURE SCOPE

The above scheme can be applied in any organization where people exercise their authorities in an online manner not on paper. This is a scheme to be applied in an environment where electronic documents are mostly produced in every process and also, when people exchange their document outside the organization for doing contracts, paying payments etc. In the later case the policy database need to be maintained at both end but with policies made in such a way that does not expose the sensitive organization details, we can consider this case as an extension of the above scheme.

## ACKNOWLEDGMENT

We thank Dr I. J. Kumar, Principal, Bharati Vidyapeeth's college of Engineering, New Delhi for his encouragement and support in carrying out the work.

## IMPLEMENTATION AND RESULTS

For
Case 1:
$\lambda_n = 6$

$\mu_n = 2$

The output shows the probabilities of person in the system. And average time of the system.

Case 2:
$\lambda_n = 8$

$\mu_n = 3$

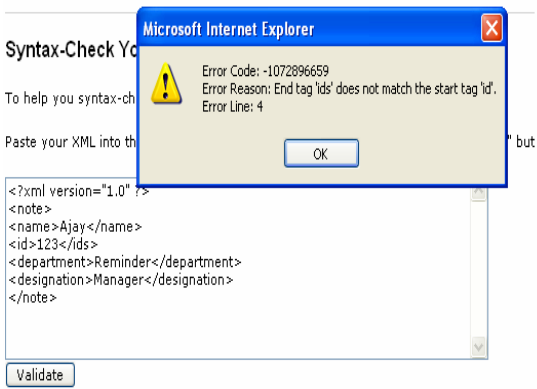

The below Snapshot shows the for XML Policy verifications

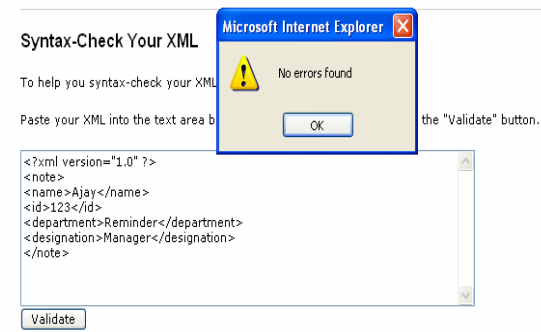

**Snapshot 1: Checking the Syntax of XML Code**

**Snapshot 2: Checking the Syntax of XML Code**

## AUTHORS PROFILE

**M.N.Doja** is a professor in Computer Science and engineering Department, Jamia Millia Islamia, New Delhi, India. he has been the Head of Department and Chairperson for research and development board for the same department, for several year.

**Dharmender Saini** received his B.Tech. from T.I.T&S in Computer Science in 1999 and M.Tech.in 2006  in Computer science and engineering from Guru jhambheswar university, hissar. During 2000-2007, he stayed in Bharati Vidyapeeth College of Engineering as Lecturer and Assistant Professor, Presently persuing PhD from Jamia Millia Islamia University,New Delhi,India